%% file: sphere_precoding_arXiv.tex
\documentclass[conference]{IEEEtran}
\input{input.tex}

\usepackage{epstopdf}
\usepackage{enumerate}
\usepackage{amsmath}
\usepackage{float}
\usepackage{color}
\usepackage{makeidx}
\usepackage{bm}
\usepackage{cleveref}
\usepackage{url}
\usepackage{steinmetz}
\usepackage{varwidth}
\usepackage{soul}
\graphicspath{{figures/}}
\usepackage{booktabs}
\usepackage{comment}
\usepackage{bbm}
\usepackage{subfigure}

\let\oldFootnote\footnote
\newcommand\nextToken\relax
\renewcommand\footnote[1]{%
	\oldFootnote{#1}\futurelet\nextToken\isFootnote}
\newcommand\isFootnote{%
	\ifx\footnote\nextToken\textsuperscript{,}\fi}

\newcommand{\sref}[1]{{Section}~\ref{#1}}

\def \rm {\mathrm}

\usepackage{algorithm}
\makeatletter
\newcommand{\algorithmname}{\ALG@name}
\renewcommand{\floatc@ruled}[2]{{\@fs@cfont #1:} #2\par}
\makeatother
\usepackage[noEnd=false]{algpseudocodex}
\tikzset{algpxIndentLine/.style={draw=black}}
\algrenewcommand{\alglinenumber}[1]{\bfseries\footnotesize #1}
\algrenewcommand{\textproc}{}
\algrenewcommand{\algorithmicrequire}{\textbf{Input:}}  
\algrenewcommand{\algorithmicensure}{\textbf{Output:}} 

\begin{document}

\title{Sphere Precoding for Robust Near-Field Communications}

\IEEEoverridecommandlockouts  

\author{Hao~Luo,~Yu Zhang,~and~Ahmed~Alkhateeb \thanks{The authors are with the Wireless Intelligence Lab in the School of Electrical, Computer, and Energy Engineering, Arizona State University, email: \{h.luo, y.zhang, alkhateeb\}@asu.edu.}}

\maketitle

\begin{abstract}	
	Near-field communication with large antenna arrays promises significant beamforming and multiplexing gains. 
	These communication links, however, are very sensitive to user mobility as any small change in the user position may suddenly drop the signal power. This leads to critical challenges for the robustness of  these near-field communication systems. 
	In this paper, we propose \textit{sphere precoding}, which is a robust precoding design to address user mobility in near-field communications.
	To gain insights into the spatial correlation of near-field channels, we extend the one-ring channel model to what we call one-sphere channel model and derive the channel covariance considering user mobility.
	Based on the one-sphere channel model, a robust precoding design problem is defined to optimize the minimum signal-to-interference-plus-noise ratio (SINR) satisfaction probability among mobile users.
	By utilizing the eigen structure of channel covariance, we further design a relaxed convex problem to approximate the solution of the original non-convex problem.
	The low-complexity solution effectively shapes a sphere that maintains the signal power for the target user and also nulls its interference within spheres around the other users. 
	Simulation results highlight the efficacy of the proposed solution in achieving robust precoding yet high achievable rates in near-field communication systems.
\end{abstract}

\input{Ch1_Introduction.tex}
\input{Ch2_system_model.tex}
\input{Ch3_problem.tex}
\input{Ch4_solutions.tex}
\input{Ch5_results.tex}
\input{Ch6_conclusion.tex} 

\bibliographystyle{IEEEtran}

\end{document}

%% file: input.tex
\usepackage{amsfonts}
\usepackage{times}
\usepackage{graphicx}
\usepackage{latexsym}
\usepackage{dsfont}
\usepackage{amssymb}
\usepackage{amsmath}
\usepackage{cite}
\usepackage{verbatim}

\newtheorem{algorithm}{Algorithm}

\newcommand{\figref}[1]{{Fig.}~\ref{#1}}

\def\bb0{{\mathbb{0}}}

\def\ba{{\mathbf{a}}}
\def\bb{{\mathbf{b}}}

\def\bff{{\mathbf{f}}}
\def\bg{{\mathbf{g}}}
\def\bh{{\mathbf{h}}}

\def\bp{{\mathbf{p}}}
\def\bq{{\mathbf{q}}}

\def\bw{{\mathbf{w}}}
\def\bx{{\mathbf{x}}}

\def\b0{{\mathbf{0}}}

\def\bA{{\mathbf{A}}}
\def\bB{{\mathbf{B}}}

\def\bF{{\mathbf{F}}}

\def\bI{{\mathbf{I}}}

\def\bR{{\mathbf{R}}}

\def\bU{{\mathbf{U}}}

\def\bbC{{\mathbb{C}}}

\def\bbE{{\mathbb{E}}}

\def\bbR{{\mathbb{R}}}

\def\cA{\mathcal{A}}

\def\cN{\mathcal{N}}

\def\cZ{\mathcal{Z}}

\def\sf0{{\mathsf{0}}}



%% file: Ch1_Introduction.tex
\section{Introduction} \label{sec:intro}
Massive multiple-input multiple-output (MIMO) has been recognized as a promising technique to enhance beamforming gain and spatial multiplexing in the current and future wireless communication systems.
As the size of antenna array aperture increases, however, the communication system is more likely to transition from the far-field to the near-field regime.
In conventional far-field communications, the electromagnetic wavefront can be approximately modeled based on the assumption of planar wave propagation.
Under this assumption, the signal can be steered towards a specific angular direction.
However, the planar wave assumption no longer holds in near-field communications.
Instead, near-field channel modeling works with the spherical wave assumption, which introduces the distance information.
As a result, near-field operations provide the capability to precisely focus a beam onto a specific spatial position.
In practice, this property brings new challenges in precoding design since both beam focusing and interference nulling become more sensitive to user mobility.
This motivates developing a robust precoding design that can mitigate the impact of user mobility in near-field communications.

\textit{Prior work}: 
The precoding design problem for near-field communications has attracted increasing research interests due to the emergence of a new degree of freedom within the distance domain.
In~\cite{Zhang2022}, the authors study the near-field precoding design in the massive MIMO systems with different antenna configurations, including fully-digital, hybrid, and dynamic metasurface antenna architectures.
Furthermore, the precoding design for tri-polarized holographic MIMO surface is investigated in~\cite{Wei2023}.
In~\cite{GC2023}, the authors tackle the joint problem of user scheduling and precoding design in MIMO systems with extremely large aperture arrays, considering imperfect channel state information.
To address user mobility with robust precoding design, previous studies have mainly focused on far-field operations.
For instance, in \cite{Lu2022,Wang2022,Lin2022}, the authors employ a posterior channel model to characterize channel uncertainties under different mobility assumptions.
The precoding design is formulated as an expected sum-rate maximization problem, which can be solved using several optimization techniques.
These far-field approaches, however, can not be directly applied to the near-field due to the assumption of planar wave propagation.

In this paper, we develop a robust precoding design, referred to as \textit{sphere precoding}, for near-field communications.
The contributions of this paper can be summarized as follows:
\begin{itemize}
    \item To determine the spatial correlation of near-field channels, we propose a one-sphere channel model for near-field operation and derive the channel covariance that accounts for user mobility.
    \item Based on the one-sphere channel model, we formulate a robust precoding design problem that accounts for user mobility and ensures users experience stable receive SINR during movement.
    \item Because of the non-convex nature of the original problem, we design a relaxed convex problem to achieve efficient and tractable solution by leveraging the eigenstructure of the channel covariance.
\end{itemize}
The simulation results demonstrate that the proposed sphere precoding can achieve robust performance in near-field communications by adaptively design the beam focusing and interference nulling according to user mobility.

%% file: Ch2_system_model.tex
\section{System and Channel Models} \label{sec:sys}
\begin{figure}[t!]
	\centering
	
	\includegraphics[width=0.489\textwidth]{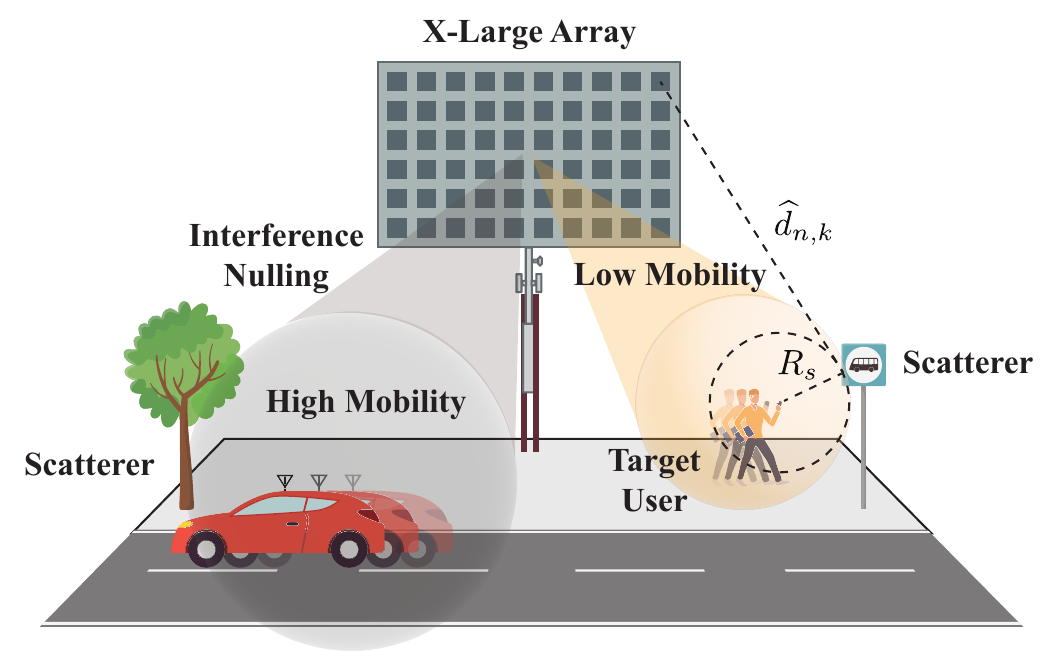}
	
	\caption{This figure illustrates the considered near-field scenario, where a base station is equipped with an extremely large array to serve mobile users. The precoding design depends on users speed and surrounding scatterers, which provides both beam focusing and interference nulling with robustness.}	
	\label{fig:sys_model}
\end{figure}

In this section, we present system and channel models in the adopted near-field communication system.
An illustration of the system is shown in  \figref{fig:sys_model}.

\subsection{System Model}
We consider a massive MIMO system, where a base station communicates with $K$ single-antenna users.
The base station is at a height of $H_\rm{BS}$ and equipped with an extremely large uniform planar array (UPA) containing $N = N_\rm{V} \times N_\rm{H}$ antenna elements.
$N_\rm{V}$ and $N_\rm{H}$ denote the numbers of the antennas in the vertical and horizontal dimensions, respectively.
We focus on the reactive near-field region, i.e., the communication distances fall within the range shorter than the Fraunhofer distance but longer than the Fresnel distance \cite{Liu2023}.
Let $\bh_k$ represent the downlink channel vector.
The signal received by the $k^\rm{th}$ user is given by
\begin{equation} \label{eq:received_signal}
    y_k = \underbrace{\bh_{k}^{H} \bff_k s_k}_{\text{Desired signal}} + \underbrace{\sum_{l \neq k}^{} \bh_{k}^{H} \bff_l s_l}_{\text{Inter-user interference}} + n_k,
\end{equation}
where $\bF = \left[ \bff_1, \ldots, \bff_K \right] \in \bbC^{N \times K}$ is the precoding matrix, and $\left[ s_1, \ldots, s_K \right]^T \in \bbC^K$ denotes the transmitted data symbols.
$n_k \sim \cN(0, \sigma^2)$ is the receive noise at the $k^{th}$ user.
Besides, we assume that the users have mobility during the data transmission.
Specifically, each mobile user is characterized by a random velocity $u_k$ and a random moving direction defined by the azimuth and elevation angles $\theta_k$, $\phi_k$, relative to the initial position $\bq_k = \left[ q_{k}^x, q_{k}^{y}, q_{k}^{z} \right]^T$.
After the $k^{\rm{th}}$ user moves for a time duration of $\Delta t$, the coordinate vector of the new position can be expressed as
\begin{align}
    \widehat{\bq}_{k}(\Delta x_k, \theta_k, \phi_k) = \bq_k + \begin{bmatrix}
        \Delta x_k \sin\phi_k \cos\theta_k \\
        \Delta x_k \sin\phi_k \sin\theta_k \\
        \Delta x_k \cos\phi_k
      \end{bmatrix},
\end{align}
where $\Delta x_k = u_k \Delta t$ is the moving distance of the $k^{\rm{th}}$ user.

\subsection{One-Sphere Channel Model}
To determine the spatial fading correlation of the near-field channel $\bh_k$, we propose the one-sphere channel model by extending the one-ring model~\cite{Shiu2000}. 
In particular, we (i) adopt the near-field channel model instead of the far-field channel model, and (ii) account for the local scattering in the spherical space surrounding the receiver.
In this scenario, scatterers are assumed to be uniformly distributed within the spherical area.
Without loss of generality, we assume the antenna array is placed on the y-z plane.
The coordinate vector of the $n^\rm{th}$ antenna element $\bp_n \in \bbR^3$ can be defined as
\begin{multline}
    \bp_{n} = \left[ 0, h_n-\frac{N_\rm{H}-1}{2}, H_{\rm{BS}}+v_n-\frac{N_\rm{V}-1}{2} \right]^{T} \cdot d,
\end{multline}
where $n = h_n + N_\rm{H} v_n + 1$, $h_n \in \{ 0, \ldots, N_\rm{H}-1 \}$, $v_n \in \{ 0, \ldots, N_\rm{V}-1 \}$, i.e., the element in the $v_n^{\rm{th}}$ row and $h_n^{\rm{th}}$ column of the antenna array, and $d$ is the spacing between antennas.
Assuming static users, we can express the coordinate vector of the scatterer surrounding the $k^\rm{th}$ user as 
\begin{align}
    \bg_{k}(r_s, \theta_s, \phi_s) = \bq_k + \begin{bmatrix}
        r_s \sin\phi_s \cos\theta_s \\
        r_s \sin\phi_s \sin\theta_s \\
        r_s \cos\phi_s
      \end{bmatrix},
\end{align}
where $r_s$ is the distance between the user and the scatterer.
$\theta_s$ and $\phi_s$ are the azimuth and elevation angles of the scatterer, relative to the user.
Then, the distance between the $n^{\rm{th}}$ antenna element and the scatterer of the $k^\rm{th}$ user can be calculated as
\begin{equation}
    d_{n,k}(r_s, \theta_s, \phi_s) = \| \bp_n - \bg_{k}(r_s, \theta_s, \phi_s) \|_2.
\end{equation}
Let $\lambda$ denote the wavelength of the carrier frequency, and $R_s$ represent the radius of the spherical region containing the scatterers.
The channel covariance $\bR_k = \bbE\left[\bh_k \bh_k^H\right]$ can be defined as \eqref{eq:cov_static}.
\begin{figure*}[t!]
    \begin{equation} \label{eq:cov_static}
        \left[ \bR_{k} \right]_{m,n} =  \frac{1}{2\pi^2 R_s} \int_{0}^{R_s} \int_{0}^{2\pi} \int_{0}^{\pi} \exp \{ -j \frac{2\pi}{\lambda} [ d_{m, k}(r, \theta, \phi) - d_{n, k}(r, \theta, \phi) ] \} d r d \theta d\phi.
    \end{equation}
\end{figure*}
For mobile users, the distance between the $n^{\rm{th}}$ antenna element and the scatterer of the $k^\rm{th}$ user can be calculated as
\begin{multline} \label{eq:mobile_distance}
    \widehat{d}_{n,k}(\Delta x_k, \theta_k, \phi_k, r_s, \theta_s, \phi_s) \\
    = \| \bp_n - \widehat{\bq}_{k}(\Delta x_k, \theta_k, \phi_k) - 
    \begin{bmatrix}
        r_s \sin\phi_s \cos\theta_s \\
        r_s \sin\phi_s \sin\theta_s \\
        r_s \cos\phi_s
    \end{bmatrix} \|_2.
\end{multline}
If the position and velocity of the user are estimated, we can establish a spherical transmission zone representing the potential locations of the user and scatterers during movement.
Mathematically, the transmission zone can be expressed as
\begin{equation} \label{eq:trans_zone}
    \boldsymbol{\cZ}_k = \left\{ \bx \in \bbR^3 \ | \ \| \bx - \bq_k \|_2 \leq \Delta x_k + R_s  \right\}
\end{equation}
Then, based on~\eqref{eq:mobile_distance} and~\eqref{eq:trans_zone}, we can define the channel covariance matrix over the transmission zone as presented in \eqref{eq:cov_mobile}, which can be further approximated by~\eqref{eq:cov_mobile_approx}.
\begin{figure*}[t!]
    \small{
    \begin{align} 
        & \left[ \bR_{k}\right]_{m,n} \nonumber \\
        &=  \frac{1}{4 \pi^4 \Delta x_k R_s} \int_{0}^{\Delta x_k} \int_{0}^{2\pi} \int_{0}^{\pi} \int_{0}^{R_s} \int_{0}^{2\pi} \int_{0}^{\pi} \exp \{ -j \frac{2\pi}{\lambda} [ \widehat{d}_{m, k}(x, \theta_k, \phi_k, r, \theta_s, \phi_s) - \widehat{d}_{n,k}(x, \theta_k, \phi_k, r, \theta_s, \phi_s) ] \} dx d\theta_k d\phi_k dr d\theta_s  d\phi_s \label{eq:cov_mobile} \\
        &\approx \frac{1}{2 \pi^2 (\Delta x_k + R_s)} \int_{0}^{\Delta x_k + R_s} \int_{0}^{2\pi} \int_{0}^{\pi} \exp \{ -j \frac{2\pi}{\lambda} [ d_{m, k}(r, \theta, \phi) - d_{n, k}(r, \theta, \phi) ] \} dr d\theta d\phi. \label{eq:cov_mobile_approx}
    \end{align}}
    \hrule
\end{figure*}
Let the singular value decomposition of the channel covariance be expressed as $\bR_k = \bU_k \boldsymbol{\Lambda}_k \bU_k^H$.
The channel $\bh_k$ can be formulated using the Karhunen-Lo\`eve representation, given by
\begin{equation}
    \bh_k = \bU_k (\boldsymbol{\Lambda}_k)^{1/2} \bw_k,
\end{equation}
where $\bw_k \sim \cN(0, \bI) \in \bbC^{\rm{rank}(\bR_k)}$.

%% file: Ch3_problem.tex
\section{Problem Definition} \label{sec:prob}
In this paper, our primary goal is to develop a robust precoding method for near-field communication, where mobile users experience consistent and stable SINR performance even as they move.
To be more specific, we aim to maximize the minimum SINR satisfaction probability among the mobile users.
With the received signal given by~\eqref{eq:received_signal}, the SINR can be written as
\begin{equation}
    \rm{SINR}_k = \frac{| \bh_k^H \bff_k |^2}{\sum_{l \neq k}^{}| \bh_k^H \bff_l |^2 + \sigma^2}.
\end{equation}
Then, the optimization problem can be expressed as
\begin{gather} \label{eq:opt_prob}
	\begin{aligned}
		\max_{\bF} \ \min_{k} \quad & \Pr_{\bh_k}\{ \rm{SINR}_k  \geq \gamma_k \} \\
		\textrm{s.t.} \quad & \| \left[ \bF \right]_{:,k} \|_2 \leq 1, \ \forall k \in \left\{ 1, \ldots, K \right\},
	\end{aligned}
\end{gather}
where $\gamma_k$ is the target SINR value of the $k^\rm{th}$ user.
To design a robust precoding approach for near-field communication, however, we need to overcome the following challenges.
First, in the near field, the signal strength of the beamforming becomes highly concentrated at a focal point as the number of antenna increases, which is vulnerable to the variations in user position.
Likewise, in terms of interference, it can be hard to nullify the signals of non-target users, especially when considering user mobility.
Therefore, conventional precoding techniques, e.g., zero-forcing precoding, prove ineffective within this scenario.
In the next section, we introduce our proposed solution to address these challenges.

%% file: Ch4_solutions.tex
\section{Proposed Solution} \label{sec:sol}
In this section, we first present the key idea of the proposed robust precoding design.
Then, we delve into the details of the developed sphere precoding method.
Finally, we discuss the practical issues of the sphere precoding.

\subsection{Key Idea}
Given that \eqref{eq:opt_prob} is non-convex and has no close-form solution, we propose an optimization approach to provide an approximation.
Specifically, assuming the users positions and velocities are estimated, we can construct a spherical transmission zone~\eqref{eq:trans_zone} for each user, which covers the potential future user and scatterer locations.
Then, we design a precoding vector for each user that achieves two objectives:
(i) A near-constant level of received power is provided among the transmission zone (sphere) of the target user;
(ii) Interference nullification in the transmission zones (spheres) around the other users.
While our approach involves an approximation, in \sref{sec:sim}, we will show the effectiveness of our solution in terms of robustness against user mobility.
%

\subsection{Sphere Precoding}
Assuming the channel covariance matrices~\eqref{eq:cov_mobile_approx} of users transmission zones are known, we first extract the singular vectors from the channel covariance.
This is motivated by the fact that the eigenstructure of the channel covariance reveals which spatial directions are statistically associated with the signal components of the channel.
Then, for the precoding matrix $\bF$, we repeatedly solve an optimization problem to obtain the precoding vector of each user.
To provide a user with consistent signal power during movement, we propose to perform an equal projection onto the singular vectors corresponding to the transmission zone of the target user. 
This objective can be formally expressed as
\begin{equation}
	\bU_k^H \bff_k = C\boldsymbol{1},
\end{equation}
where $C$ is a normalization constant.
To address the inter-user interference, we aim to achieve zero projection onto the singular vectors that correspond to the transmission zones of the non-target users.
To that end, we adopt an interference constraint, which can be written as
\begin{equation}
	\| \bU_l^H \bff_k \|_2 \leq \epsilon,\ \forall l \neq k,
\end{equation}
where $\epsilon$ is the tolerance constant of the interference.
With the definitions above, the sphere precoding vector of an user can be obtained by solving the following optimization problem
\begin{gather}  \label{eq:opt_sphere_precoding}
    \begin{aligned}
        \min_{\bff_k} \quad & \| \bU_k^H \bff_k - C\boldsymbol{1} \|_2 \\
        \textrm{s.t.} \quad & \| \bU_l^H \bff_k \|_2 \leq \epsilon,\ \forall l \neq k, \\
        \quad & \| \bff_k \|_2 \leq 1.
    \end{aligned}
\end{gather}
It is worth noting that, by removing the interference constraint, \eqref{eq:opt_sphere_precoding} is reduced to a single-user beamforming problem that considers resilience to user mobility.
The detailed process of sphere precoding is provided in Algorithm \ref{alg1}.

\begin{algorithm}[t]
	\caption{Sphere Precoding}
	\label{alg1}
	\begin{algorithmic}[1]
		
		\Require \begin{varwidth}[t]{\columnwidth} 
			Channel covariance matrix $\bR_k$.
		\end{varwidth}
		
		\Ensure Precoding matrix $\bF$. 

		\For{$k=1$ \textbf{to}  $K$} \!\!\!\!  

		\State Compute the singular vectors $\bU_k$ of the channel covariance by using singular vector decomposition, $\bR_k = \bU_k \boldsymbol{\Lambda}_k \bU_k^H$
		
		\EndFor
		
		\For{$k=1$ \textbf{to}  $K$} \!\!\!\!  

		\State Calculate the precoding vector $\bff_k$ by solving \eqref{eq:opt_sphere_precoding}.

		\State $[\bF]_{:, k} = \bff_k$
		
		\EndFor
		
	\end{algorithmic}
\end{algorithm}

\subsection{Practical Operation}
In this subsection, we discuss the potential operations of sphere precoding in practice.
Specifically, the sphere precoding relies on the knowledge of users positions and velocities to define the spherical regions and construct the channel covariance matrices.
This information, for instance, can be acquired by leveraging radar sensing aided communication \cite{Luo23}, where radar sensors are deployed at the base station to provide observations regarding the service area.
The radar sensory data contains range, velocity, and angular information of mobile users, which can be utilized in the sphere precoding design. This information could also be obtained leveraging other multi-modal sensors like vision, LiDAR, or just position sensors.
Furthermore, it is noteworthy that our proposed solution provides robustness against estimation errors of position and velocity.
This resilience stems from employing solid spherical regions to ensure consistent communication performance for mobile users.
Even when these spheres may shift, enlarge, or reduce in size due to estimation errors, the sphere precoding is able to maintain a certain level of performance under such circumstances. 
Therefore, the potential practicality of sphere precoding appears promising.

%% file: Ch5_results.tex
\section{Simulation Results} \label{sec:sim}
In this section, we evaluate the performance of the proposed sphere precoding with numerical simulations.

\begin{figure}[!t]
	\centering
	
	\subfigure[Zero-forcing precoding]{
		\includegraphics[width=0.48\textwidth]{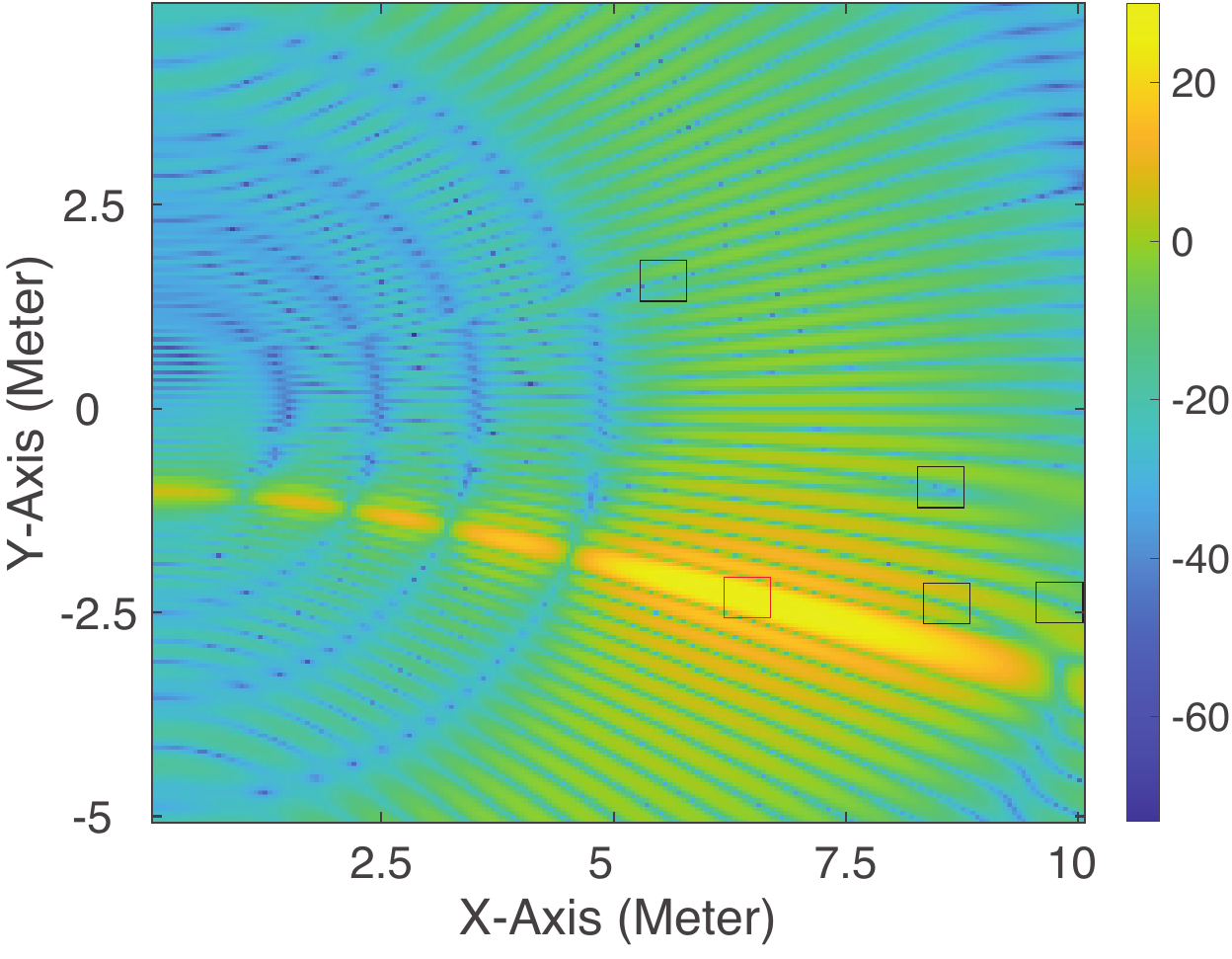}
		\label{fig:beam_zf}
	}
	\hfill 
	\subfigure[Sphere precoding]{
		\includegraphics[width=0.48\textwidth]{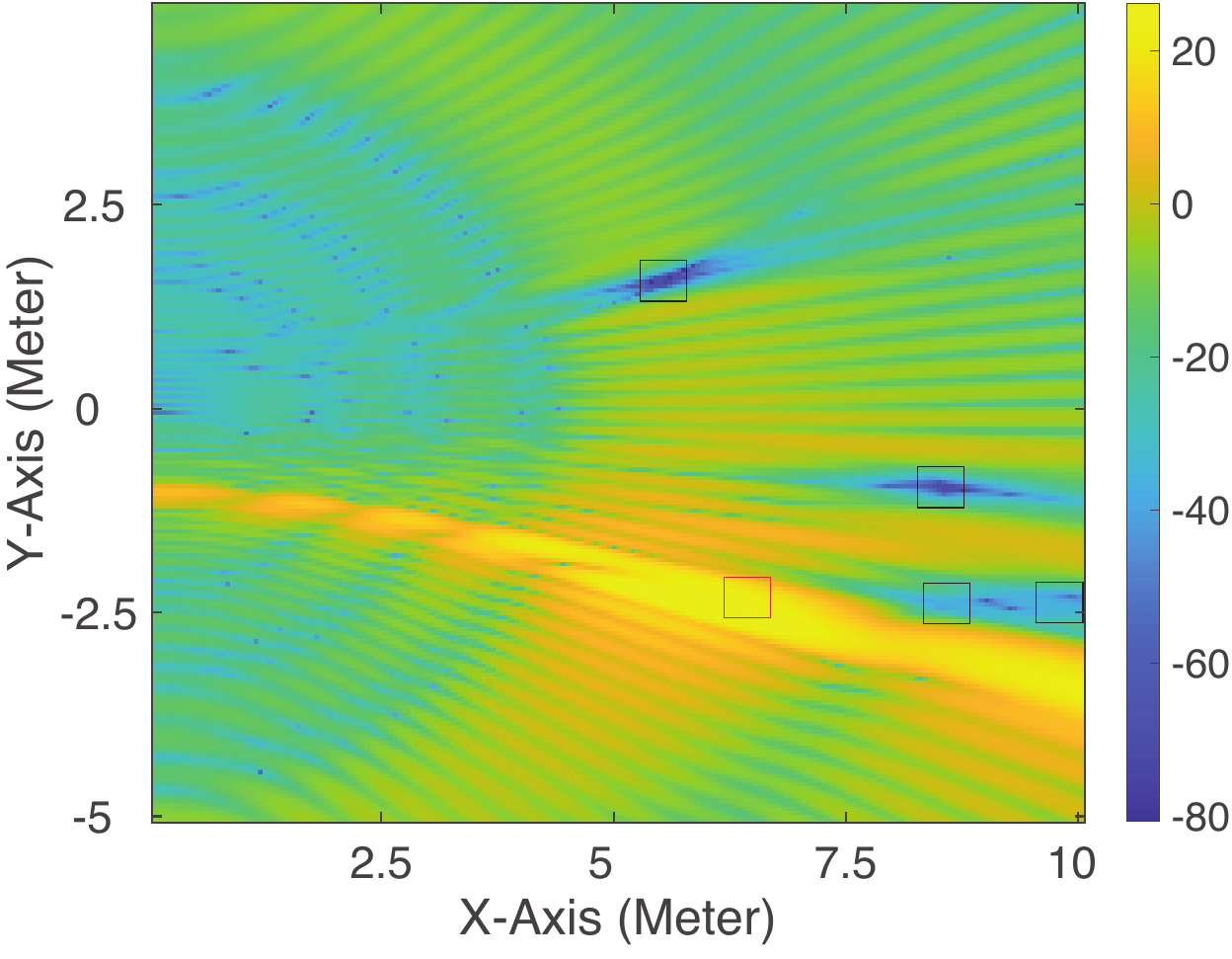}
		\label{fig:beam_sphere}
	}
	
	\caption{This figure shows a comparison between the beamforming patterns of zero-forcing and sphere precoding. The base station is at the origin,  the target user is marked by a red square, and the other (non-target) users are marked by black square. User mobility is assumed to be 0.056 m.  In Figure (b), sphere precoding effectively focuses/nullifies beam power around the spherical region of target/non-target users. This leads to more robust communication performance against user mobility.}	
	\label{fig:beam_pattern}
	\vspace{-10pt}
\end{figure}

\subsection{Simulation Setup}
\textbf{System Model:}
We consider the base station is equipped with a $16\times64$ UPA, i.e., $N_\rm{V}=16$ and $N_\rm{H}=64$, at a height $H_\rm{BS}=3$m.
The system operates at a carrier frequency $28$ GHz, and the spacing between antennas equals $\lambda/2$.
We randomly drop $K=5$ mobile users in the near-field region, and the received noise power is set to $\sigma^2=10^{-2}$.

\textbf{One-sphere Channel Model:}
The radius of the spherical region containing the scatterers is set to $R_s=0.05$m.
Given the mobility assumption of the users, i.e., moving distance, we can define the transmission zones based on~\eqref{eq:trans_zone}.
To construct the channel covariance matrix~\eqref{eq:cov_mobile_approx}, we uniformly generate $100$ channel samples in each transmission zone.
For channel realization, the number of scatterers surrounding each user is set to $10$.

\textbf{Sphere Precoding:}
In the optimization problem~\eqref{eq:opt_sphere_precoding}, the normalization constant is set to $C=1$, and the tolerance constant of interference equals $\epsilon=10^{-3}$.
To solve this problem, we adopt the solver MOSEK~\cite{mosek} via CVX~\cite{cvx}.

\textbf{Benchmark:}
We compare the proposed sphere precoding with the following methods:
(i) \textit{Conjugate Beamforming}: the LoS channel between the base station and the initial position is used as the precoding vector for each user;
(ii) \textit{Dominant Eigenvector}: the dominant eigenvector of the channel covariance is considered as the precoding vector for each user;
(iii) \textit{Equal Projection}: the inter-user interference constraint is removed from~\eqref{eq:opt_sphere_precoding};
(iv) \textit{Zero-Forcing Precoding}: the LoS channels corresponding to the users initial positions are utilized;
(v) \textit{Fractional Programming}\cite{Lin2022}: the precoding design objective is to maximize the expected sum-rate.
Note that the first three methods do not address the inter-user interference.
Also, we compute an upperbound by assuming the perfect knowledge of instantaneous channel information and with no inter-user interference.
In the simulation, we conduct $100$ independent experiments with different user distributions to obtain the results.

\begin{figure}[!t]
	\centering
	
	\subfigure[Average SINR]{
		\includegraphics[width=0.45\textwidth]{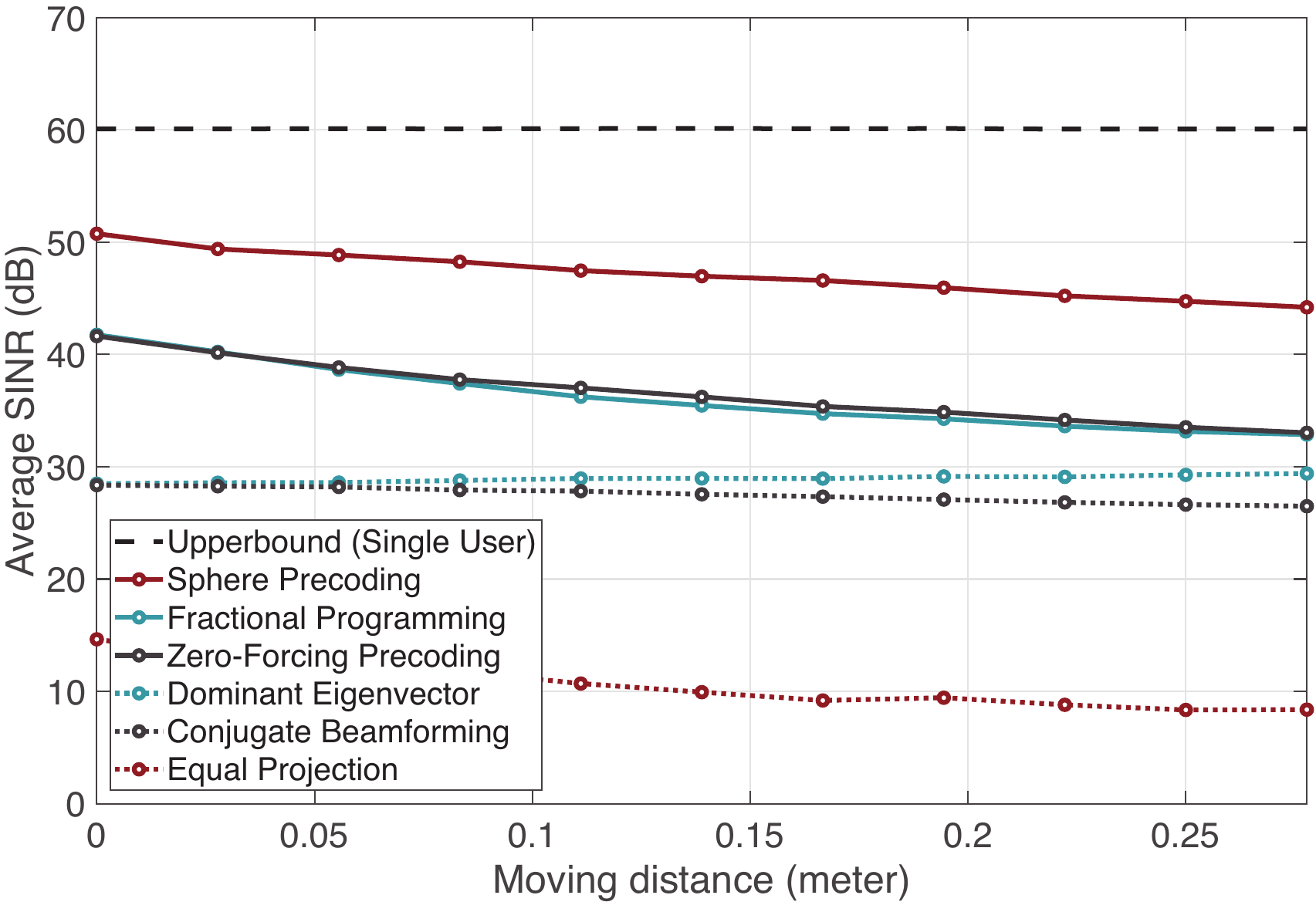}
		\label{fig:avg_sinr}
	}
	\hfill 
	\subfigure[CDF]{
		\includegraphics[width=0.45\textwidth]{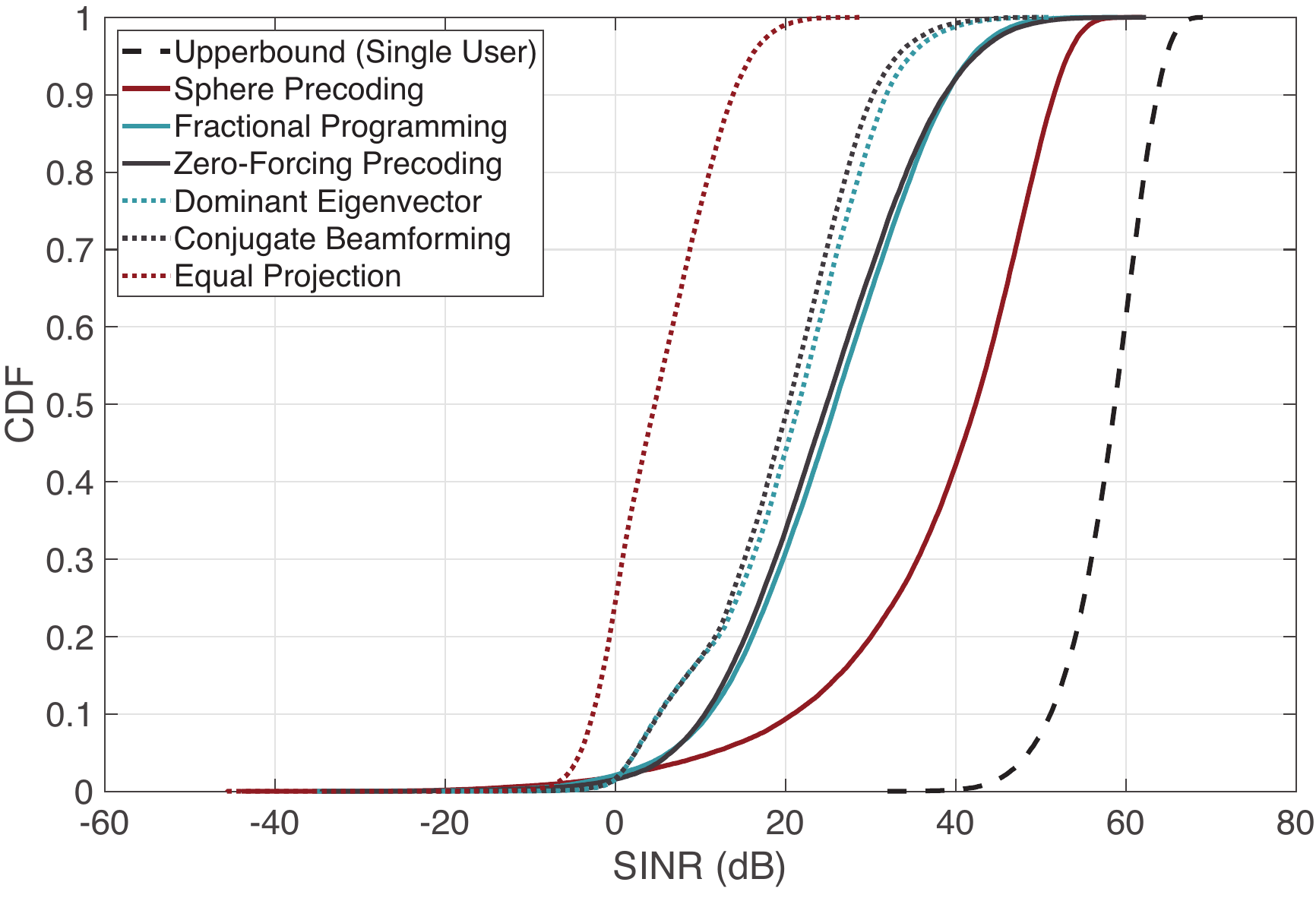}
		\label{fig:cdf_sinr}
	}
	
	\caption{The evaluation of the proposed solution. In (a), the average SINR of the proposed solution is compared with the benchmark methods with varying mobility assumption, i.e., moving distance. In (b), the CDF of SINR at the moving distance of $0.139$ m is presented.}	
	\label{}
\end{figure}

\subsection{Performance Evaluation}
In \figref{fig:beam_pattern}, the beamforming patterns of precoding are visualized.
As depicted in \figref{fig:beam_zf}, \textit{Zero-Forcing Precoding} is able to effectively concentrate beam power and nullify the interference regarding the users initial positions.
When users have mobility, however, the communication performance tends to decline because of the outdated channel information. 
By contrast, although the propose sphere precoding sacrifices peak power at the target user, it is able to construct null areas for non-target users according to their mobility, as shown in \figref{fig:beam_sphere}.
Consequently, the proposed solution exhibits robustness in terms of user mobility.

In \figref{fig:avg_sinr}, the average SINR of the proposed solution is evaluated with varying mobility assumption.
It can be observed that the methods that disregard inter-user interference suffer from performance degradation.
To be more specific, the \textit{Equal Projection} method is worse than \textit{Conjugate Beamforming} and \textit{Dominant Eigenvector} due to the distribution of the signal power across the transmission zone.
Thanks to the zero projection onto the transmission zones of non-target users, the proposed sphere precoding outperforms other methods under different mobility assumption.
In \figref{fig:cdf_sinr}, the cumulative distribution function (CDF) regarding the receive SINR is presented.
Specifically, the proposed solution shows better robustness against user mobility compared to other approaches.
For instance, sphere precoding achieves around $90\%$ of samples with SINR exceeding $20$ dB, while \textit{Zero-Forcing Precoding} and \textit{Fractional Programming} attain only $70\%$.
This underscores that sphere precoding provides users with more consistent communication performance during their movement.

%% file: Ch6_conclusion.tex
\section{Conclusion} \label{sec:conclusion}
In this paper, we develop a robust precoding design to address the impact of user mobility in near-field communications.
To investigate the spatial correlation of the near-field channels, we establish a one-sphere channel model and derive the channel covariance within a spherical transmission zone, which is defined based on user mobility.
With the one-sphere channel model, we propose sphere precoding, which leverages the singular vectors of channel covariance to incorporate user mobility into precoding design. 
The proposed solution ensures stable received power for the target user while effectively mitigating inter-user interference in mobile scenarios.
Simulation results demonstrate the effectiveness of the proposed sphere precoding in achieving robust near-field communications.